\documentstyle[preprint,aps]{revtex}
\begin{document}
\draft
\title{Closed Abrikosov Vortices in a Superconducting Cylinder
}
\author{V. A. Kozlov and  A. V. Samokhvalov}
\address{Institute of Applied Physics, Russian Academy of Sciences\\
46 Uljanov Street, 603600 Nizhny Novgorod, Russia
}
\maketitle
\begin{abstract}
The new type of solutions of the London equation for type-II
superconductors is obtained to describe the ring-shaped (toroidal)
Abrikosov vortices. The specific feature of these solutions is the
self-consistent localization of both the supercurrent and the
magnetic field, enabling one to construct compact magnetic structures
inside a superconductor. The torus vortex contraction caused by
the vortex instability leads to the destruction of the Cooper
pairing and the formation of a normal electron stream in the
vicinity of the torus axis. The thermodynamic condition for the
excitation of a small closed vortex by a bunch of charged
particles contains the fine-structure constant as a determining
parameter.
\end{abstract}
\widetext
\section{Introduction}
\label{sec:intro}

A magnetic flux penetrates into type-II superconductors (SC)
as lines of Abrikosov vortices having one magnetic flux quantum
\cite{abrik1}. These vortices and their interaction with
inhomogeneities and
defects of the material (pinning) determine both the magnetic
properties of type-II SC and the ability of SC to carry the
superconducting current \cite{james2}. In the vortex core,
where the magnetic field
achieves its maximum, superconductivity is destroyed, and the
modulus of the complex order parameter , $\psi$ , describing
the superconducting properties of the material, goes to zero. Central
points of the vortex core where  $\vert \psi \vert = 0$
define the location of
vortex line , (VL), while the phase of  $\psi$  changes by a multiple
of $2\pi$ in going round the loop encircling this line. The latter
circumstance leads to the flux quantization in the vortex, so that
a solitary vortex with a rectilinear VL contains a single flux
quantum, $\Phi_0  = \pi \hbar c/e$.

To describe the behavior of Abrikosov vortices in type-II SC
with small coherence length compared to the magnetic field
penetration depth  $\lambda$  , it is convenient to apply
the London model involving the principle of superposition for
currents and fields \cite{abrik3,shmidt4}.
In the London approximation, the vortex
is determined by
the VL location as an external parameter. The vortex lines in
type-II superconductors reproduce the structure of the magnetic
field lines . Thus, the homogeneous magnetic field , ${\bf H}_0$ ,
generates
a two-dimensional lattice of rectilinear VLs \cite{abrik1,hueben5}.
More complex
vortex structures are possible when the magnetic field is produced
by an external current flowing through the SC. In the case of a SC
cylinder with a bias current, the magnetic field lines represent a
set of concentric circles and any VL is a circle with its center
lying at the cylinder axis \cite{campb6,ullmai7}.The current-induced
magnetic
structures in cylindrical SC wires have been
discussed in \cite{andr8,andr9}.
The continuously collapsing magnetic structures in cylindrical
wires have been considered in \cite{gorter10,gorter11,koppe12}
to explain the SC resistive state.

The present paper deals with the new type of solutions of the
London equation. These solutions describe the ring-shaped (or
toroidal) Abrikosov vortices in a type-II SC. The VL of such a
vortex is a circle of a finite radius.The toroidal vortex arising
in SC cylinders exhibit some peculiarities compared to the Abrikosov
vortex with a rectilinear VL. In Sec.\ \ref{sec:2}, the
magnetic field,
the magnetic flux and the free energy of a solitary closed toroidal
vortex in a unbounded SC are calculated. In Sec.\ \ref{sec:3},
the toroidal solution of the London equation for a superconducting
cylinder for arbitrary relations between the London penetration
depth, the cylinder radius and the toroidal vortex size is derived,
and the vortex stability is discussed . In Sec.\ \ref{sec:4}, the
excitation of toroidal vortices by a bunch of charged particles is
considered.

\section{A\ Closed\ Abrikosov\ Vortex in an\ Unbounded\
type-II\ Superconductor}
\label{sec:2}

     The London approach to superconductors is appropriate for
the case $\lambda\gg\xi$~, where $\lambda$ is the London penetration
depth and  $\xi$
is the coherence length. This approximation is expected to be
justified for new high-T  superconductors. To find the structure of a
solitary closed Abrikosov vortex, we consider the solution of the
London equation with a closed VL consisting of a circle of radius,
$R_s$~, inside an unbounded SC. In the London model, the magnetic
field distribution in the vortex , ${ \bf H}({ \bf r})$~,
at distances  $r$  from the
center of the normal core, is yielded by the
London equation \cite{abrik3}:
\begin{equation}
{\bf H}+\lambda^2 rotrot{\bf H}
 = \Phi_0{\bf e}_v\delta({\bf r}-{\bf r}_s).
\end{equation}
Here ${ \bf r}_s$ is the radius-vector defining the VL location
in space
and  ${ \bf e}_v$   is the unit vector directed along the VL
(Fig.~\ref{fig1}). In the
cylindrical coordinate system  $(r,\varphi,z)$~, whose  $z = 0$
plane coincides
with the VL plane, the magnetic field , ${ \bf H}({ \bf r})$~,
has only an
azimuthal component , $H(r,z)$~, and the basic equation (1)
can be written in the form:
\begin{equation}
{\partial^2 h_v \over \partial \rho^2} + {1\over \rho }
{\partial h_v \over \partial \rho}
+ { \partial^2 h_v \over \partial \zeta^2}
 - (1+ {1 \over \rho^2}) h_v
 = - \delta(\rho-\rho_s) \delta(\zeta).
\end{equation}
Here  $\rho=r/ \lambda$~, $\zeta=z/ \lambda$~, $\rho_s=R_s/ \lambda$~, and
$h_v(\rho,\zeta) = H(r,z) \lambda^2/ \Phi_0$   is
the dimensionless magnetic field. To solve Eq.(2) the Fourier-Bessel
transform has been used and the magnetic field,
$h_v(\rho,\zeta)$~, has been represented as follows:
\begin{mathletters}
\begin{equation}
h_v(\rho,\zeta) = {1 \over \sqrt{2 \pi}}
{ \int_{-\infty}^{+\infty} dp \exp {(ip \zeta)} }
{ \int_0^{+\infty} dq f_v(p,q) q J_1( \rho q) } ,
\end{equation}
\begin{equation}
f_v(p,q) ={1 \over \sqrt{2 \pi}}
\int_{-\infty}^{+\infty} {d \zeta} \exp (-ip \zeta)
\int_0^{+\infty} {d \rho} h_v( \rho, \zeta) \rho J_1( \rho q) ,
\end{equation}
\end{mathletters}
where $J_1$ is the Bessel function of the first kind. The differential
equation (2) for the Fourier-Bessel components , $f_v$~, expressed in
the spectral variables  $(p,q)$~,  is reduced to the simple algebraic
relation:
\begin{equation}
f_v(p,q)(1+p^2+q^2) = \rho_s J_1(\rho_s q) / \sqrt{2 \pi}.
\end{equation}
Using Eq.(3.a) one can return to the initial coordinates
$( \rho, \zeta)$
and obtain the magnetic field distribution ,$h_v$~, in a solitary
closed Abrikosov vortex:
\begin{equation}
h_v(\rho,\zeta) ={ \rho_s \over 2}
\int_0^{+\infty} dq { q J_1(\rho q) J_1(\rho_s q)
\exp( - \vert \zeta \vert \sqrt{1+q^2} )
\over \sqrt{1+q^2} } .
\end{equation}
Expression (5) has a logarithmic divergence at the VL center
$(\rho=\rho_s, \zeta=0)$ , because the London equation (1)
is not valid in
the region of the normal vortex core, where
${\vert {\bf r}-{\bf r}_s \vert} \leq \xi$. The
magnetic field was thus truncated in the usual way at distances  $\xi$
from the VL center \cite{abrik3}. The toroidal-shaped vortex with a large
radius, $R_s \gg \lambda$~, is similar to the linear vortex, and the magnetic
field in this vortex varies significantly over distances of the
order of $\lambda$. In the opposite limit  $\xi \ll R_s \ll \lambda$~,
the spatial region
occupied by the toroidal vortex is determined primarily by the
radius , $R_s$ .

The London equation (1) implies the following expression for
the Gibbs free energy , $G$ :
\begin{equation}
G = {\lambda^2 \over {8 \pi}}
\int_S d{\bf S} \Bigl[ {\bf H} rot{\bf H} \Bigr]
+ {1 \over {8 \pi}} \int_V dV {{\bf H}
\Bigl( {\bf H}+\lambda^2 rot rot {\bf H} \Bigr) } .
\end{equation}
For the considered solitary toroidal vortex, the surface integral
is zero and the free energy , $G_v$  , is determined by:
\begin{equation}
G_v( \rho_s) = {\Phi_0^2 \over {8 \pi \lambda^2}}
L_s h_s( \rho_s)  ,
\end{equation}
where  $h_s( \rho_s) \equiv h_v( \rho= \rho_s, \zeta=0)$
is the magnetic field value at the
VL center and  $L_s$  is the VL length. In case  $R_s \gg \lambda$,
the free
energy of the toroidal vortex coincides with the energy , $G_L$ , of
the rectilinear Abrikosov vortex of the length , $L_s$ \cite{abrik3}:
\begin{equation}
G_v \simeq G_L = {\Bigl( \frac{ \Phi_0}{4 \pi \lambda} \Bigr)}^2
L_s {K_0( \xi / \lambda)},
\xi \ll \lambda \ll R_s,
\end{equation}
where  $K_0$  is the McDonald function. Figure~\ref{fig2} shows the field ,
$h_s$~,
and the energy , $G_v$~,  versus the radius , $R_s$ .
Since the energy ,
$G_v$~, grows monotonically with $R_s$~, the toroidal vortex
is unstable,
tending to contract toward the  $z$  axis and ultimately collapsing.

By integrating the distribution (5) over the  $( \rho, \zeta)$
half-plane, we can calculate the magnetic flux , $\Phi_v$~,
in the annular vortex:
\begin{equation}
\Phi_v( \rho_s) = \Phi_0 \Bigl( 1- \rho_s {K_1( \rho_s)} \Bigr),
\end{equation}
where  $K_1$ is the modified Bessel function of the second kind.
Therefore, the flux , $\Phi_v$~, depends on the VL radius , $R_v$~,
and tends to the asymptotic value
$\Phi_0$  at  $R_s \gg \lambda$ ( Fig.~\ref{fig2}).
Utilizing the expression
\begin{equation}
{\bf j} = \Big( \hbar e/ m \Bigr) {\vert \psi \vert}^2
\Bigl( \nabla \chi - (2e/ \hbar c) {\bf A} \Bigr)
\end{equation}
for the superconducting current density , ${\bf j}$ \cite{abrik3}~,
one can write
down the condition for the fluxoid of the closed vortex:
\begin{equation}
\Phi_v + {2 \pi \lambda^3 \over c}
\int_{-\infty}^{+\infty} {d \zeta} {j_z( \rho=0, \zeta)} = \Phi_0 .
\end{equation}
Here  $\psi = {\vert \psi \vert} \exp(i \chi)$  is the complex
order parameter in the
Ginzburg-Landau theory , ${\bf A}$ is the vector potential of
the magnetic field, and $j_z( \rho=0, \zeta)$ is the  $z$
component of the current
density , ${\bf j}$~, yielded by expression (10).
Using Eqs. (9,11) and
assuming that the relation
\begin{equation}
\int_{-\infty}^{+\infty} {d \zeta} {j_z( \rho=0, \zeta)}
\simeq 2R_s \overline{j_z} / \lambda
\end{equation}
is valid for a small VL circle of radius , $R_s \ll \lambda$,
one can estimate
the average current density , $ \overline{j_z}$ ,
at the  $z$  axis :
\begin{equation}
\overline{j_z} \simeq {c \Phi_0 \over {4 \pi \lambda^2 R_s} }.
\end{equation}
A decrease of a closed vortex of a radius , $R_s$ ,
is thus accompanied
by an increase in the superconducting current density at the  $z$
axis. One can easily obtain that for the VL radii , $R_s \leq R_s^d$ ,
where
\begin{equation}
R_s^d = 3 \sqrt{3} \pi \xi ,
\end{equation}
the average current density , $\overline{j_z}$,
exceeds the current density of
the superconductivity destruction ,$j_c$ \cite{abrik3}:
\begin{equation}
j_c = {c H_{cm} \over {3 \sqrt{6} \pi \lambda} }
= {c \Phi_0 \over {12 \sqrt{3} \pi^2 \lambda^2 \xi}} .
\end{equation}
Note that  $R_s^d \gg \xi$, so the London approximation is still valid.
If the average current density ,$\overline{j_z}$,
exceeds the  depairing current
density , $j_c$ , then the Cooper pairing is destructed and a normal
electron stream forms in the vicinity of the  $z$  axis.The closed
vortex contraction finishes in confluence of the additional normal
region occurring at the  $z$  axis with a normal region existing in
the vortex core. The phase difference of the order parameter,
present in going round VL, disappears due to collapse and induces
thereby the voltage , $V$ , yielded by the relation ,
$\partial \varphi / \partial t = 2eV / \hbar$.
This voltage leads to the vortex energy dissipation. The processes
taking place near the torus axis during the closed vortex collapse,
resemble the formation of the phase-slip centers in narrow
superconducting channels (whiskers), when the current density in these
channels exceeds the critical value \cite{ivlev13}.
Since the topology of
the magnetic field in the toroidal vortex is determined by the
external current flowing through the SC, one can assume that the
periodic occurrence and collapse of these vortices create the
resistive state in the superconducting channels with a bias current.
This is valid when the transverse dimensions of the channel are
larger than $\xi$ and the intrinsic magnetic field of the current
affects essentially the transverse structure of the solution
\cite{hueben5,campb6}.

\section{A Closed Abrikosov Vortex in a Superconducting cylinder of
an Arbitrary Radius}
\label{sec:3}

     Let us consider the influence of the SC boundary on the
structure and the properties of a closed vortex. In a superconducting
cylinder of an arbitrary radius , $R_c$  (Fig.~\ref{fig1}), the dimensionless
field distribution , $h( \rho, \zeta) = H(r,z) \lambda^2 / \Phi_0$~,
in a toroidal vortex with
a VL of radius , $R_s \leq R_c$~, is described by the following
equation :
\begin{equation}
{\partial^2 h \over \partial \rho^2} + {1\over \rho }
{\partial h \over \partial \rho}
+ { \partial^2 h \over \partial \zeta^2}
- (1+ {1 \over \rho^2}) h
 = - \delta(\rho-\rho_s) \delta(\zeta).
\end{equation}
which coincides with Eq.(2). The corresponding boundary condition
for the magnetic field produced by the vortex on the surface of
the superconducting cylinder, is :
\begin{equation}
h( \rho_c=R_c/ \lambda, \zeta) = 0 .
\end{equation}
Therefore, to determine the structure of the closed toroidal
Abrikosov vortex, one should solve Eq.(16) with the boundary
condition (17) in the region  $\rho \leq \rho_c$.
Due to linearity of (16,17)
we represent their solution as a superposition of the solutions
for the solitary closed vortex (5) and for the homogeneous
boundary problem inside the cylinder:
\begin{equation}
{\partial^2 h_c \over \partial \rho^2} + {1\over \rho }
{\partial h_c \over \partial \rho}
+ { \partial^2 h_c \over \partial \zeta^2}
- (1+ {1 \over \rho^2}) h_c = 0,
\end{equation}
%
\begin{equation}
h_c( \rho= \rho_c, \zeta) = -h_v( \rho= \rho_c, \zeta),
\rho \leq \rho_c .
\end{equation}
By representing the solution in the form
\begin{equation}
h( \rho, \zeta) = h_v( \rho, \zeta) + h_c( \rho, \zeta),
\end{equation}
we generalize the well-known procedure of determining the structure
of the Abrikosov vortex with a rectilinear VL parallel to the plane
surface of the SC, by means of supplementing the vortex with its
mirror image , $h_c( \rho, \zeta)$.
The field and the current produced by the
image are directed opposite to the corresponding values of the
primary vortex, $h_v( \rho, \zeta)$ \cite{abrik3}.
In this case the image is distorted
by the curvilinear surface of the cylinder. The solution of the
homogeneous equation (18) for the cylinder  $ \rho \leq \rho_c$  can be
written as :
\begin{equation}
h_c(\rho,\zeta) = {1 \over \sqrt{2 \pi}}
{ \int_{-\infty}^{+\infty} dp \exp {(ip \zeta)} }
C(p) I_1( \rho \sqrt{1+p^2} ) ,
\end{equation}
where $I_1$ is the modified Bessel function of the first kind. The
integration constant , $C(p)$ , is given by the boundary condition
(19):
\begin{equation}
C(p) = - \frac{ \rho_s}{ \sqrt{2 \pi} }
\frac{ I_1( \rho_s \sqrt{1+p^2} ) K_1( \rho_c \sqrt{1+p^2} ) }
{ I_1( \rho_c \sqrt{1+p^2} ) } .
\end{equation}
The solution of the homogeneous boundary problem (18,19), valid
for  $\vert \rho_c - \rho_s \vert \geq \xi/ \lambda$ , is :
\begin{equation}
h_c( \rho, \zeta) = - \frac{ \rho_s}{ \pi}
\int_0^{+\infty} dp \cos(p \zeta)
\frac{ I_1( \rho \sqrt{1+p^2} ) I_1( \rho_s \sqrt{1+p^2} )
K_1( \rho_c \sqrt{1+p^2} ) }
{I_1( \rho_c \sqrt{1+p^2} ) } .
\end{equation}
Thus, the magnetic field structure , $h( \rho, \zeta)$ ,
of the solitary closed
Abrikosov vortex inside a superconducting cylinder of an arbitrary
radius , $R_c$ , is fully determined by the relations (5,20,23). An
example of the magnetic field structure in the closed vortex,
computed by (5,20,23), is given in Figure~\ref{fig3}. It is essential that
the cylinder surface affects the vortex structure only in the
surface layer of a thickness  $\lambda$.
It is convenient to represent the
magnetic flux , $\Phi$ , in this vortex as:
\begin{equation}
\Phi = \Phi_v( \rho_s)
- \Phi_v( \rho_c) \frac{\rho_s I_1( \rho_s)}
{\rho_c I_1( \rho_c)}.
\end{equation}
Here the magnetic flux , $\Phi_v$ , in a solitary toroidal vortex,
 determined by (9), is expressed explicitly. According to the boundary
condition (17), the magnetic field produced by the vortex, vanishes
on the surface and exists only inside the superconductor cylinder.
The Gibbs free energy in this case is given by the formula
\begin{equation}
G( \rho_s) = \frac{ \Phi_0^2 }{8 \pi \lambda^2}
L_s h_s( \rho_s),
\end{equation}
where  $h_s( \rho_s)$ is the magnetic field in the core,
$h_s( \rho_s) = h_s( \rho= \rho_s, \zeta=0)$,
with allowance for the surface influence:
\begin{equation}
h_s( \rho_s) = h_v( \rho_s,0) + h_c( \rho_s,0) .
\end{equation}
The magnetic flux , $\Phi( \rho_s)$~, and the free energy ,
$G( \rho_s)$~, versus the
position of the toroidal vortex inside the cylinder are shown in
Fig.~\ref{fig4}. It can be readily seen that the closed vortex in the
cylinder remains unstable and the interaction with the surface
does not stabilize the vortex. Depending on the radius of the
created VL, the vortex either moves towards the cylinder axis or
disappears on the surface.

\section{The excitation of a closed toroidal vortex by a charged
particle bunch}
\label{sec:4}
     Now we discuss the possibility for exciting a closed toroidal
vortex by an external bunch of charged particles. The azimuthal
structure which is required for forming a toroidal vortex is
offered by the self-magnetic field , $H_q$~, of a charge , $q$~,
moving at a velocity , $V$~, parallel to the  $z$  axis [14]:
\begin{equation}
H_q = \frac{q \beta \gamma}{\lambda^2}
\frac{ \rho}{ (\rho^2+ \gamma^2 \zeta^2)^{3/2} }.
\end{equation}
where  $\beta = V/c$~, $c$ is the velocity of light and
$\gamma = (1 - \beta^2 )^{-1/2}$
is the relativistic factor. Instead of dealing with the rigorous
nonstationary problem, we use the following thermodynamic
condition:
\begin{equation}
G_q = G_v - \frac{1}{4 \pi} \int d^3r (H_q H) \leq 0 ,
\end{equation}
which estimates the vortex formation energy in the external
magnetic field , $H_q$. Inequality  (28) imposes an upper limit ,
$R_s^m$ \@ $(R_s \leq R_s^m )$~, on the radius of the vortex
line which can be excited in
this fashion. For a charge in relativistic motion  $( \beta \simeq 1)$~,
the dimensionless maximum radius , $ \rho_s^m  = R_s^m / \lambda$~,
satisfies the equation:
\begin{equation}
\frac{2}{ \pi h_s( \rho_s^m)}
\frac{1- \exp(- \rho_s^m)}{ \rho_s^m} = \frac{ \alpha^{-1}}{Z} .
\end{equation}
Here  $\alpha = e^2 / \hbar c$
is the fine-structure constant (spin-orbit
constant),  $Z = q/e$  and  $e$  is the charge of an electron.
Therefore, the only physical constant which determines the excitation
of a toroidal vortex, as a macroscopic entity, is the fine-structure
constant , $\alpha$ . Figure~\ref{fig5} plots the maximum vortex radius ,
$\rho_s^m$~,
yielded by Eq.(29) versus the magnitude of the charge , $q$ .It
should be mentioned, however, that the London approximation used
here, is valid only when $R_s^m \gg \xi$. Hence a toroidal
vortex can be
excited only by a rather large magnitude of the moving charge,
i.e.\  $Z = q/e \simeq \alpha^{-1} \simeq 137$ .
This may be realized, for example,
by a bunch of charged particles. Note, that the closed vortex
excitation by a bunch of charged particles is ,in a sense, the
inverse process of the vortex collapse described in Sec.\ \ref{sec:2}
of the present paper.

\section{Conclusion}
\label{concl}
     The magnetic field structure, the magnetic flux and the free
energy of a closed Abrikosov vortex with a toroidal structure of
a VL in an unbounded SC and inside a superconducting cylinder of
an arbitrary radius, have been calculated in the London
approximation. The magnetic flux and the free energy
of the toroidal vortex
strongly depend on the VL radius. The individual toroidal vortex
is always unstable: it either contracts towards the torus axis or
emerges on the cylinder surface. The VL contraction (decrease of
its radius) is accompanied by breaking of the Cooper pairs and
generates a normal electron stream in the vicinity of the torus
axis. The possibility of exciting a closed toroidal vortex by an
external bunch of charged particles has been analyzed. It has been
shown that the fine-structure constant is the determining parameter
of the thermodynamic condition for the excitation of a vortex
by a moving bunch of charged particles.

\begin{figure}
\caption{Toroidal Abrikosov vortex inside a superconducting cylinder
of radius ,$R_c$ .  VL (circumference of radius $R_s$ ) is shown by a
dash-dotted line.}
\label{fig1}
\end{figure}

\begin{figure}
\caption{1 - Magnetic field at the center of the vortex line, $h_s$ ;
2 - the magnetic flux, $\Phi_v$   ; inset - the free energy , $G_v$,
all as functions of the radius of the vortex line, $R_s$. Shown for
comparison by the dashed line is the free energy of a linear
Abrikosov vortex with a length , $L_s = 2 \pi R_s$
$( \kappa = \lambda / \xi = 100 )$.}
\label{fig2}
\end{figure}

\begin{figure}
\caption{Magnetic field structure of a closed toroidal Abrikosov
vortex inside a superconducting cylinder  ( $R_s = 3 \lambda$;
$R_c = 5 \lambda$; $ \kappa = \lambda / \xi = 100$ ).}
\label{fig3}
\end{figure}

\begin{figure}
\caption{1 - Magnetic flux , $\Phi$ ; 2 - free energy , $G$, of a closed
toroidal Abrikosov vortex inside a superconducting cylinder versus
the vortex line radius, $R_s$
( $R_c = 5 \lambda$; $ \kappa = \lambda / \xi = 100$ ).}
\label{fig4}
\end{figure}

\begin{figure}
\caption{Maximum permissible radius of a vortex line,
$R_s^m = \lambda \rho_s^m$,
versus the magnitude of the charge of moving particles, $q$
( $ \kappa = \lambda / \xi = 100$ ).}
\label{fig5}
\end{figure}


\begin{references}
\bibitem{abrik1} A.A.Abrikosov,\@ Zh.\ Eksp.\ Teor.\ Fiz.\ { \bf32}
(1957) 1442\@ [Sov.\ Phys.\ JETP\ { \bf5} (1957) 1174].
\bibitem{james2} D.Saint-James, G.Sarma and E.J.Tomas,\@ Type-II
Superconductivity (Pergamon Press, Oxford, 1969).
\bibitem{abrik3} A.A.Abrikosov,\@ Fundamentals\ of the\ Theory of\
Metals\@ (North-Holland,\@ Amsterdam,\@ 1988 ).
\bibitem{shmidt4} V.V.Shmidt and G.S.Mkrtchan,\@ Usp.\ Fiz.\ Nauk\ { \bf112}
(1974) 459.
\bibitem{hueben5} R.P.Huebener,\@ Magnetic Flux Structures in Superconductors
(Springer-Verlag, Berlin, 1979).
\bibitem{campb6} A.M.Campbell and J.E.Evetts,\@ Critical
Currents in Superconductors\@ (Barnes \& Noble, New-York, 1972)
\bibitem{ullmai7} H.Ullmaier,\@ Irreversible Properties of
Type-II Superconductors\@ (Springer-Verlag, Berlin, 1975).
\bibitem{andr8} A.F.Andreev and Yu.V.Sharvin,\@ Zh.\ Eksp.\ Teor.\ Fiz.\
{ \bf53} (1967) 1499\@ [Sov.\ Phys.\ JETP\ { \bf26} (1967) 865].
\bibitem{andr9} A.F.Andreev,\@ Zh.\ Eksp.\ Teor.\ Fiz.\ { \bf54}
(1968) 1510\@ [Sov.\ Phys.\ JETP\ { \bf27} (1968) 809].
\bibitem{gorter10} C.J.Gorter,\@ Physica\ { \bf23} (1957) 45.
\bibitem{gorter11} C.J.Gorter and M.L.Potters,\@  Physica\
{ \bf24} (1958) 169.
\bibitem{koppe12} H.Koppe,\@ Phys.\ Stat.\ Sol.\ { \bf17} (1966) K229.
\bibitem{ivlev13} B.I.Ivlev and N.B.Kopnin,\@ Usp.\ Fiz.\ Nauk\ { \bf142}
(1984) 435.
\bibitem{landau14} L.D.Landau and E.M.Lifshitz,\@ The Classical
Theory of Fields\@ (Nauka, Moscow, 1988).
\end{references}
\end{document}